Subreddit R/unpopularopinion:

Against the Spiral of Silence

Dasom Eom, Linh Hoang, Gerry Oei

Georgia Institute of Technology

**Abstract**

As a social animal, human conforms to the customs of society, and their behavior and opinions are greatly influenced by social norms [5]. Accordingly, people choose to hide their opinion in front of others when they feel their idea is against majority opinion. This social phenomenon is called the spiral of silence [4]. Recently, the advent of internet technology and online communication has made people to stand up and express their opinion openly in online space and to break the wall of the spiral of silence [12]. The subreddit r/unpopularopinion has been designed to allow users to share and discuss socially unpopular thought frankly. In this study, r/unpopularopinion was analyzed based on the findings from community observation and participant interviews, and we introduce a descriptive study to show how the subreddit r/unpopularopinion helps its users to get over the spiral of silence. Even though the environment of r/unpopularopinion encourages users to freely express their opinion, the dominant population of this community is young white males, and only the opinions representing them are supported by general users and brought up to the front page. Consequently, the minority opinions are neglected and it results in another spiral of silence phenomenon in the community.

*Keywords: Unpopular opinion, Online community, Spiral of silence, Reddit*

Subreddit r/unpopularopinion: Against the Spiral of Silence

The social exchange theory explains that people behave in a certain way based on the expectation that their actions on another person or organization will produce specific results in response to behavior [3]. Likewise, people act and think in accordance with social norms to avoid being stigmatized by others [15]. Above that, social interactions converge the behavior of individual members of society to a similar way within the scope of social discipline [1]. The spiral of silence theory describes that people are afraid to be isolated in the social group once they express their opinion against universal thoughts. Because people want to continue to belong to their group, they choose to remain silent instead of outspeaking when they expect a number of group members would not agree with their opinions [4]. Universal thoughts include the social norms that have been accumulated for the maintenance of human society over the history of mankind. Moreover, mainstream media has a great influence on the universal thoughts in the modern world. Therefore, people with opinions deviate from the norms set by media are discouraged from speaking their thought. As the information era has come, the development of internet technology has created virtual online places in which people with similar interests and opinions can gather and share their opinions freely without the risk to be alienated in their social groups in reality.

 Related works

Many researchers have worked on the connection between the 'spiral of silence' theory and social media. Hampton and his colleagues studied to seek the public's opinions about the Snowden leak in 2014 and their willingness to openly discuss the revelation of Snowden in various in-person and online settings [7]. Although social media creators and its proponents had originally expected that social media platforms would provide enough diverse venues in which people might freely express their opinions, the study found out that people in reality still chose to say nothing about sensitive subjects on Facebook and Twitter, especially if they expect that their friends and family members would disagree with them.

Chaudhry et al analyzed public comments left on news about race, racism, or ethnicity on the Canadian Broadcasting Corporation News Facebook page. They examined whether users holding racist viewpoints would be less likely to express their opinion that was against the majority with fear to be socially marginalized. The study concluded that a number of the vocal minority was comfortably expressing their unpopular views on Facebook, and this posed a question on the explanatory power of the spiral of silence theory [2].

While Facebook and Twitter are predominantly nonanonymous platforms, Reddit users can choose their level of anonymity. Unlike other subreddits that are centered around hobbies like games or movies, the Subreddit r/unpopularopinion is an anonymized discussion forum. Malaspina provided qualitative evidence that most of the controversial opinions about political affiliation in the Italian election were expressed in online anonymous and pseudonymous forums [13]. People tend to express their opinion more honestly in anonymous and pseudonymous places, but even in the anonymized forum, the effect of the spiral of silence still exists. In the paper by Yun and Park, it was suggested that "in an online forum where a majority opinion exists, users may decide not to post a message against the majority opinion." [18].

**Subreddit r/unpopularopinion**

The subreddit r/unpopularopinion is an online community designed for people who believe that their ideas are unpopular, and it provides them a place to freely express their opinions without the risk to be isolated from their social group. In r/unpopularopinion, users facilitate discussions on highly controversial topics, which are too sensitive to be discussed in real life. As of March 20, 2019, this subreddit r/unpopularopinion has more than 400,000 subscribers and more than 18,000 comments a day, making it the seventh most commented subreddit [17]. It is one of the fastest growing subreddits: the number of r/unpopularopinion's subscribers has increased more than 10 times in 2018.

**The Present Study**

In the present study, we analyze the phenomenon of the spiral of silence in r/unpopularopinion using Amy Jo Kim's community design principles [10]. The purpose of this study is to show what aspect of the subreddit r/unpopularopinion helps users to overcome the spiral of silence. Based on the qualitative data we collect from participant observation and interviews with active users of the community, we will examine social phenomena in r/unpopularopinion. To better understand r/unpopularopinion, we will discuss the following questions in this online community. First, what is the unpopular opinion for members of r/unpopularopinion? Second, what features of the r/unpopularopinion encourage people to break out of the spiral of silence? Third, given that voting rule works differently in r/unpopularopinion compared to the general Reddit voting system, what effects does it have on r/unpopularopinion? Fourth, is there an echo chamber in this subreddit? Finally, we will conclude the paper with the suggestions to solve the problems in r/unpopularopinion.

## Methods

In this study, a descriptive research design was used to analyze the research hypotheses on the subreddit unpopularopinion. To understand the trends and the interaction between users of r/unpopularopinion, we have observed the community for three months. Above that, we conducted interviews on regular users and moderators of the community to collect qualitative data.

**Participant observations**

To get insight into the users' behaviors and their social interactions on r/unpopularopinion, we made a participant observation on the community from January 2019 to March 2019. We wrote on our profile description that we are researchers from Georgia Tech, although we did neither reveal our identity nor post our research purposes and activities openly in the forum to minimize our influence on the user's behaviors. We participated in discussions by commenting and posting in r/unpopularopinion. As a result, we were able to get more realistic data in the natural setting. The observed phenomena and interesting points on the community were written in field notes.

**Qualitative Interviews**

Semi-structured interviews were performed with participants in the study. Interviews were conducted in either form of instant-text chat or voice chat. Before we started the interview, we provided the participants with an informed consent form and explained their rights and the purpose of the interview. The informed consent form can be found in Appendix A. Each interview lasted from 45 minutes to one hour and a half. Interview questions were promptly modified depending on the participant's community status and their responses.

**Interview Participants**

In the qualitative interview, the participants were one female and nine male adults who are older than 18 from diverse ethnic and cultural backgrounds. Seven of the participants were regular users, and three of the participants were moderators of the subreddit r/unpopularopinion; one of the moderators, Participant 20, was the founder of the community. The participants who were involved in the interviews were fluent in English, and they had an ability to understand the questions and respond back their opinions in either written or Spoken English. The people who were younger than 18 or not fluent in English were excluded from participating in the interview. The participants were recruited from the subreddit r/unpopularopinion community using convenience sampling method. We asked each community user to volunteer in the interview via Reddit chat. The participants were not paid monetarily for participating in the interview.

**Data Analysis**

All the collected interview records were stored in text form. The voice interviews were transcribed into text using transcribing services like Rev.com. For the analysis, three coauthors of this study exchanged the interview transcripts and field notes and then analyzed r/unpopularopinion community based on Amy Jo Kim's community design principles and other published literature [10]. After thorough readings on the overall document, we discussed meaningful data to interpret the social phenomena in r/unpopularopinion better.

## Results

**Participant interviews**

**Adam (Adam)**

Adam is Adam 25-year-old white male in Washington, US. He went to college but dropped out to work in the beer brewing field. He first encountered r/unpopularopinion through a link from another subreddit. He likes r/unpopularopinion because this community supports free speech. He usually likes to interact with political, social and cultural issues as well as funny and weird opinions. He has more than 20,000 karma and regularly comments on the subreddit.

**Chris**

Chris is a 24-year-old male living in the United States. He is getting into a dental school in the fall of 2019. Currently, he is living with his parents to save more money. He is an avid Reddit user. He previously had a popular Reddit account, but after one of his posts, which contained some information related to his identity, appeared in the front page, he has changed to his current Reddit account. He has more than 30,000 karma, and he is one of the top commenters on r/unpopularopinion. At first, he had been a lurker for a couple of months, but he became an active member since then for about one and a half years. He likes to spend time arguing on topics about misogyny or politics. Also, he enjoys to comment on topics that are ironic and light-hearted like "Wakanda isn't a real place". He stated that he does not post as much as he comments; His previous posts were usually about science fiction and movies.

**Participant 3**

Astro is an 18-year-old Chinese living in the United States. He moved to the US four years ago, and he is currently attending high school. He first noticed r/unpopularopinion from the main Reddit page, and he has been interested with the contents in it. A few months ago, he subscribed r/unpopularopinion and became a regular member. For the majority of the time, he enjoys reading and commenting on trivial and funny topics like "I hate pineapple on pizza". During the period we were interviewing, there were many anti-China sentiments on r/unpopularopinion because a post with a picture of the 1989 Tiananmen Square protests had been at the front page of Reddit. He posted an opinion thread, "China is not a dystopian state", which was against the anti-China sentiments. He browsed r/unpopularopinion every day, and he noticed some problems of r/unpopularopinion. One issue he mentioned is that some opinions are reposted from time to time, and the other issue is that some opinions which he perceived to be popular tend to get a lot of upvotes.

**Participant 4**

Participant 4 is a 27-year-old male social worker from Australia. He is a moderator of r/unpopularopinion. Interesting point is that he has two Reddit accounts: one for his moderation duty and the other one, which is his main account, for his general Reddit activity. With his main account, he participates in "discussions about relationships, religion, [or] ... anything to do with cheating". He said that he joined r/unpopularopinion several months ago. The first time he encountered r/unpopularopinion, he was attracted to the discussions on serious and controversial topics in r/unpopularopinion, which is very different from the general light-hearted conversations in other subreddits. As a social worker, he has worked to help criminals. He cares a lot to rehabilitate the criminals and to prepare them to go back to society. Particularly he is concerned with felony offenders who conducted heavily stigmatized crimes such as child abuse and rape. On r/unpopularopinion, there are many sympathizing opinions like "wanting to help [those criminals] reform, change their lives"; when he encounters these opinions, he feels that his experience in r/unpopularopinion is positive and insightful.

**Participant 5**

The participant 5 is a female who graduated college a few years ago, and she is currently working as an auditor in the Eastern United States. As an active Redditor, she has participated in r/unpopularopinion for 3 years as a regular member since she finds it more fun to participate in or observe the discussions in r/unpopularopinion compared to other online communities. She defined r/unpopularopinion as a socially minded opinion forum. In addition, she stated that whether it is considered unpopular or popular depends on the understanding and the context of audiences on a question asking what is an unpopular opinion. She also mentioned one of the important functions of r/unpopularopinion is that it lets users know that they are not alone thinking like that and knowing there are other people with similar opinions reassures the users that it is okay to think it in that way. As one of the disadvantages of r/unpopularopinion, she pointed out that moderators have different standards on deleting or punishing the users.

**Participant 6**

The participant 6 is a male college student living in Eastern Canada. He started to visit r/unpopularopinion after being exposed to the interesting posts on the main Reddit page. He stated that he has received an impression from this community that there are many male users complaining about a social phenomenon like reverse sexism. To be specific, he pointed out that this community is very male-dominant subreddit and it carries a lot of racist and sexist opinions in this community. As a result, it does not reflect the opinions of minorities properly. To improve r/unpopularopinion community, he claimed that opinions in regard to race and gender issues are better not be discussed in the community since a lot of them are not acceptable opinions.

He defined the trolls compared to users with the unpopular opinion as follows:

> "Trolls are those who absolutely outlandish while having nothing to the back up their opinion or throwing something out there and then not really supporting it by any sort of intelligent or even thoughtful information to back it up."

He defined r/unpopularopinion as a community in which discussions are relatively civil, but some of the comments are straight up, mean, and disrespectful because of the anonymity of the internet. Consequently, he said that "people do not think about others' feelings as much as they would as if they are having a face-to-face discussion".

One of his main points in the interview was that:

> "There has never been an argument that has been able to change my mind or made me see or perceive from their point of view, but at least I understood what it is."

Moreover, I found out that he usually upvotes on the opinions he agrees with or he likes. He also stated that a lot of similar opinions are repeatedly posted in r/unpopularopinion. To redesign the community better, he suggested to introduce the filter the posts by category or subjects or to implement the tags as like other subreddit styles.

**Participant 7**

The participant 7 is a male moderator of r/unpopularopinion from Iceland, Europe. He encountered r/unpopularopinion through other user's profile page. Originally, He was a lurker in the subreddit r/rant since he liked to hear other people's inner thoughts. However, when he found r/unpopularopinion, he could hear a more diverse range of perspectives from this community. He stated that his experiences in r/unpopularopinion changed or challenged his existing point of view a lot. He believes that the reason why people visit r/unpopularopinion is to fulfill their "morbid curiosity in hearing controversial and dark views". People are naturally interested in controversial issues, and r/unpopularopinion is full of debatable posts. He defined unpopular opinion as a viewpoint that is outside the mainstream. According to him:

> "It has to, at the bare minimum, be something a large portion of people in any given group would disagree with. That is it fundamentally, something that will cause controversy in a normal setting or is the type of thought you feel you have to hide and cannot say without being judged by society."

He also stated a similar opinion with participant 5 on a question asking the difference between popular opinion to unpopular opinion. He said that whether it is a popular opinion or not depends on how it is perceived by general society. In general, popular opinion has a positive reception, and thus people can speak it out without "fearing of judgment". In contrast, while an opinion is popular around several groups, if it is not generally accepted by society at large, then it can be considered an unpopular opinion.

On the following question on the phenomenon of r/unpopularopinion that only the popular opinions come to the front page, the participant 7 replied that it is the main reason why he volunteered to be a moderator of r/unpopularopinion. Also, he pointed out that some of the topics are repeatedly posted over and over. Most users in r/unpopularopinion are young white males, and a lot of opinions in r/unpopularopinion speak for the group of young white males. Consequently, the opinions on the front page are supported by the group but still unpopular in general. To solve this issue of representing the opinions of only the specific group of users in r/unpopularopinion, he suggested filtering the posts by category. He also claimed that they need to regulate more strongly on reposted opinions.

**Participant 8**

Participant 8 is an 18-year-old college student from the UK. He also has a part-time job and usually spends his time in front of his computer coding for his own game. He has participated in the Reddit community for about a year or so with the current account. Previously, he had an old account which was around three to four years old. As for his participation in the subreddit r/unpopularopinion, he has participated in r/unpopularopinion for four months. When he participates in r/unpopularopinion, he primarily comments on other people's posts. Despite participating only for about four months while semi-regularly browsing around the subreddit, this participant feels like r/unpopularopinion is mostly populated by "gamerbros". In his definition, "gamerbros" are people, primarily gamers, who spend their time watching YouTube reactionary channels and they think they know a lot about politics but actually only talk about things YouTubers say. He stated that he feels that the "gamerbros" are looking for echo chambers to showcase their bigoted beliefs. The participant also pointed out that the echo chambers are generally the comments left on the posts. How the participant pointed out that echo chambers do exist in the comments of posts came out as unexpected due to how generally one would expect unpopular opinions to get backlashes. However, in r/unpopularopinion, most unpopular opinion posts result in echo chambers instead.

**Participant 9**

Participant 9 is a student who has participated in the Reddit community for about three years. In the interview, he mentioned an interesting fact that how there is a fundamental conflict in choosing what makes up an unpopular opinion through a voting system that allows for popular and widely accepted opinions to be promoted by individuals who agree with the opinions but feel as though they are unique and outside of the mainstream. Moreover, the participant pointed out that because of how posts could be manually removed by the moderators, the posts in r/unpopularopinion centers around the beliefs of what the moderators consider as "unpopular" and "popular" opinion posts instead. In contrast to the comment of participant 7 that the user base of the subreddit is mostly made of young white males, Participant 9 said that he perceives a lot of posts are written from the conservative viewpoints that usually come from the old generations.

**Participant 20**

Participant 20 is a 27-year-old male living in the Midwest United States. He makes wine for work, and he is the founder and one of the moderators of the subreddit r/unpopularopinion. In the interview, he said why he made the subreddit in the first place:

> "...I noticed there was not one when I went to make a post and I thought it might be a funny idea. And then it just sat there for maybe three or four years and then it started people more and more subscribers..."

At that time, what he considered as an unpopular opinion was basically just something humorous and along with the spirit of opinions like "I hate eggs". Thus, he did not expect r/unpopularopinion would grow up to the point in which there are so many different opinions about various topics as it is. To manage and regulate the community which has grown too huge, he had to recruit a number of moderators. According to him, the duties of moderators are as follows:

> "We do not really have anything established like a rule set or anything like that, but generally we have one or two people who focus on keeping the auto-moderator running and we'll do tweaks on that and the code. And then everyone else goes through the MOD queue and makes decisions on whether to remove topics or ban them. We have a Discord server in order to discuss decisions regarding removing topics or posts or banning users that require the consensus of all moderators..."

Therefore, the duties of the moderators are divided based on specific tasks such as tweaking the code for the CSS of the subreddit or auto moderator, while the other moderators focus on making decisions on the banning of users or removal of posts. Generally, the moderators have online meetings to discuss making decisions as they require the consensus of all moderators. Contrasting this to Participant 9's opinion about how posts are centered around the beliefs of the moderators, there are doubts as to whether or not decisions made are subjective or objective, especially for the removal of posts that are considered as "popular". He points out that due to the rapidly growing subreddit, the moderators are dependant on the auto-moderator to regulate the posts as much as possible according to the rules as the moderators are not capable of checking each and every post made. Therefore the most posts' validity is checked based on the logic of the auto-moderator, and posts that require more attention will be notified to the moderators for more detailed checking.

**Design Evaluations**

The design of subreddit r/unpopularopinion was evaluated in this section based on Amy Jo Kim's nine design principles for building community [10].

**Purpose/Mission**

On the subreddit r/unpopularopinion, in the community details section, which is located at the right tab of the page, it is explicitly stated that the purpose of the subreddit r/unpopularopinion is to promote a place for users to share any unpopular opinions they have.

**Audience**

In the demographic survey conducted in r/unpopularopinion, 70.8% of users were White, and Asians were second in rank of the population. The age distribution of r/unpopularopinion was as follows. 42.1% of users identified that they are between 18 to 24, 32.2% of users identified that they are under the age of 18, and 19.3% of users were between 25 to 34. Occupation-wise, most users were either attending college (31.8%) or still at middle-school or high-school (28.6% for both). The gender distribution of r/unpopularopinion showed that the subreddit is dominated by males (76.7%). Based on the demographic survey and participant interview, it appeared that r/unpopularopinion is dominated by young white males [16].

**Visual design**

The visual design of r/unpopularopinion is evaluated in this section. Recently, r/unpopularopinion changed its design to a new style. The overall color theme was changed from grey and black tones to green. We are comparing the old design to the new design to evaluate whether the change has brought a positive influence on user activities in r/unpopularopinion.

*Community Rules Design*

The community rules are displayed at the sidebar located at the right tab of r/unpopularopinion (Figure 1). The rules specify a guideline for users to follow while participating in r/unpopularopinion. If the user breaks any rules, the post can be deleted or the user activity can be banned in r/unpopularopinion. When a user clicks the inverted caret button next to each rule, the user can see the detailed information in regard to each term (Figure 2). In contrast, in the old Reddit design, the rules tab did not have a compressed version of the design, and the details for each rule was displayed in default (Figure 3).

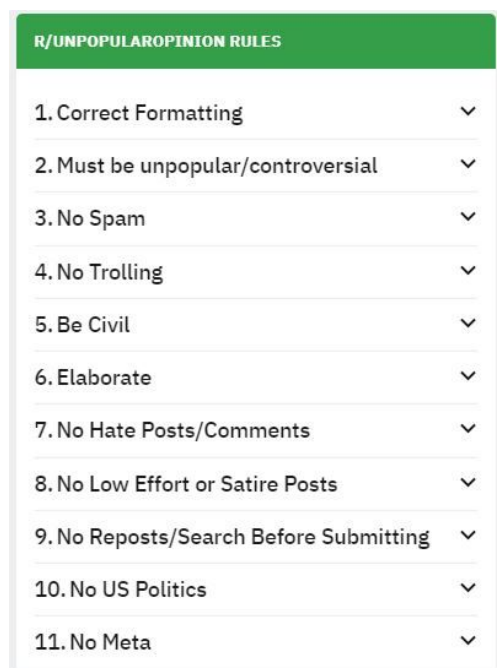

*Figure 1.* Community rules in the new Reddit design.

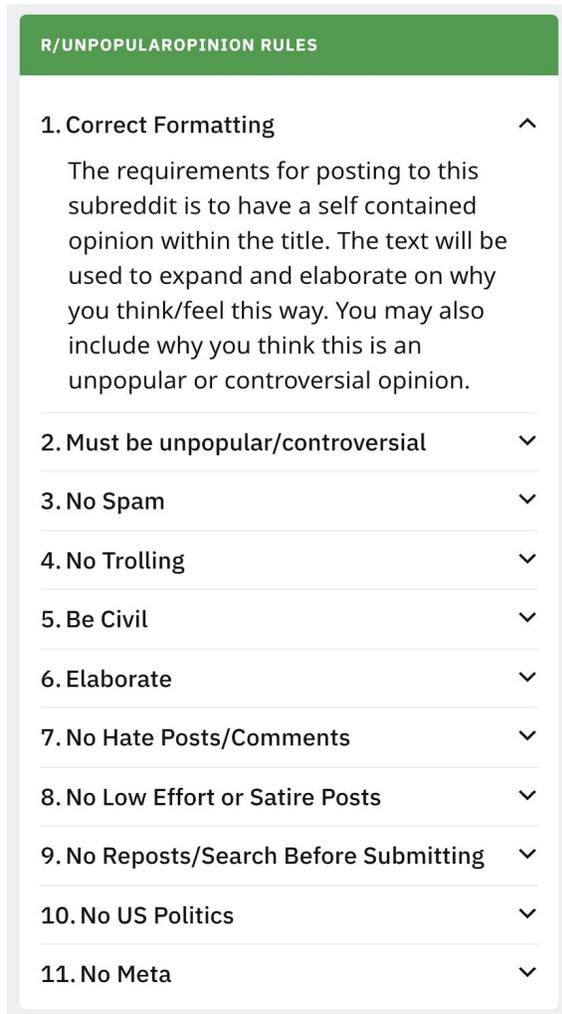
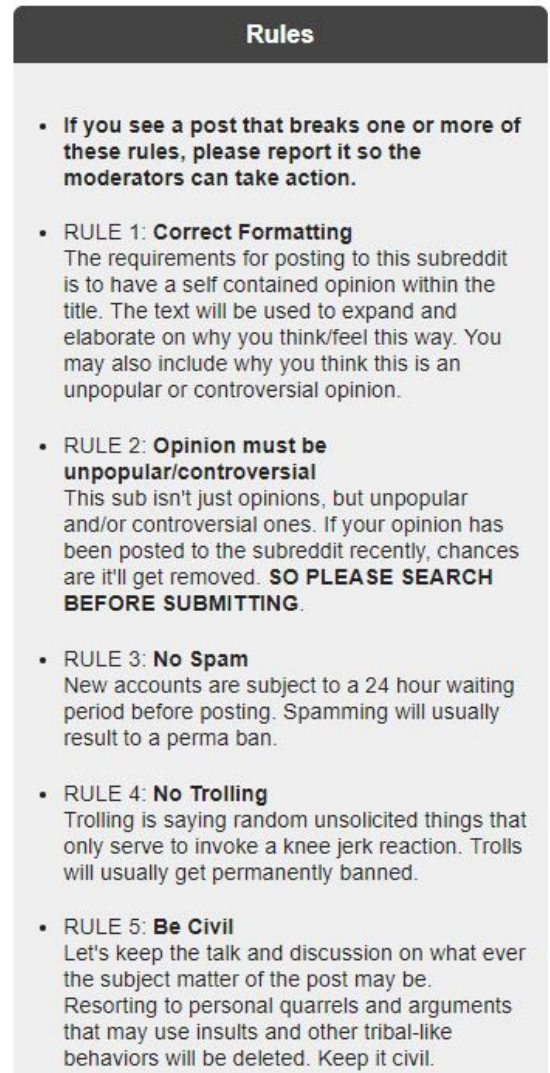

*Figure 2*. Expanding the first rule reveals the detailed explanation about the correct formatting.

*Figure 3*. Rules section in old Reddit design.

*Voting System Design*

In order to help promote "good" posts, Reddit has up and down buttons at the left side of each post as shown in figure 4. The use of these buttons is to either upvote or downvote the post. In r/unpopularopinion, this function gives users a right to regulate the community contents; Upvotes are made for posts with genuine unpopular or controversial opinions while downvotes are made for posts with popular opinions. This voting mechanism is also used for the comments left on each post as well. The number between the upvote arrow and the downvote arrow represents the total number of votes the post or the comment has received. The posts are placed in the order of voting score in r/unpopularopinion; the more score the post receives, the closer the post is put to the front page.

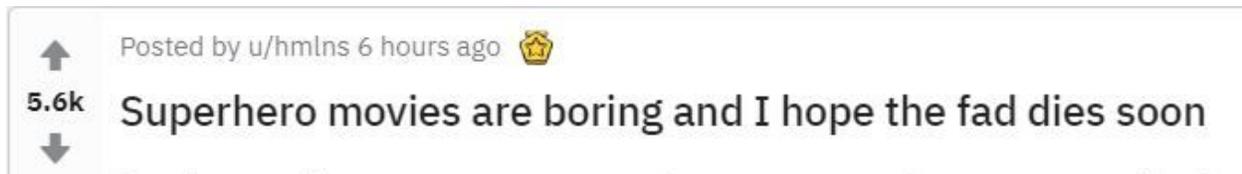

*Figure 4.* Upvote and downvote buttons at the left side of the post title.

In the old design, there existed a tooltip function in addition to the up and down buttons for the votes. This tooltip appears when users hover over the up or down buttons as seen in Figure 5 and Figure 6. The tooltip message checks if the user votes accordingly to the community rules. That is, the message notifies the users again that posts should be upvoted if they are unpopular and well-written opinions, and they should be downvoted if they are popular or badly written opinions.

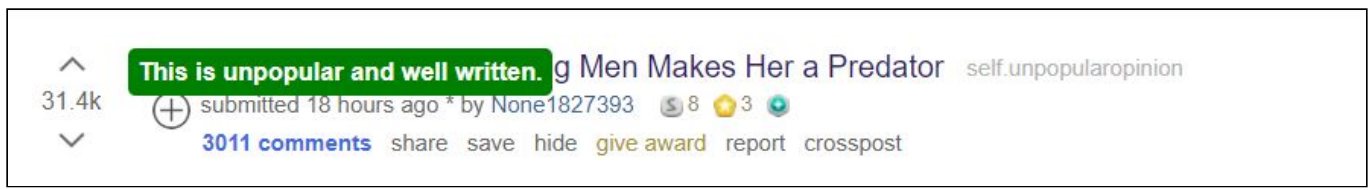

*Figure 5.* Tooltip message pops up when the user hovers over the upvote button in the old Reddit design.

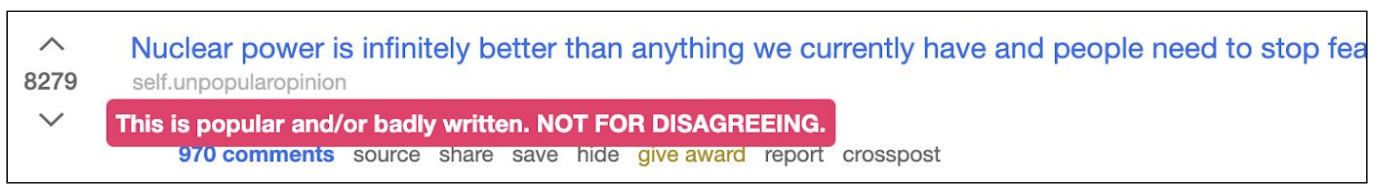

*Figure 6.* Tooltip message pops up when the user hovers over the downvote button in the old Reddit design.

*Auto-moderator Message*

When posting an opinion, bot messages pop up in the very top of the comments section of the post. The bot message is generated by the auto-moderator. The message reminds the users of the community rules again so that the users can comply with the community regulations. It ensures the users which posts to upvote and which posts to downvote and informs them political opinions posted outside the politics megathread will be deleted (Figure 7).

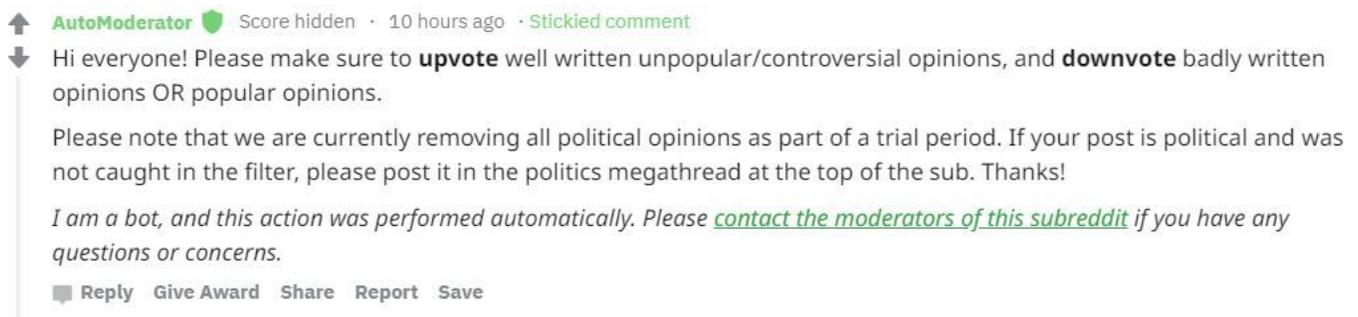

*Figure 7.* Auto-moderator comment

*Flair*

In the old Reddit design, the feature of 'flair' had existed to help to filter out contents in r/unpopularopinion. The 'flair' is a tag that can be attached to posts. Previously in the subreddit r/unpopularopinion, the flairs had been used to indicate which posts are good unpopular opinion posts. In Figure 8, the tag of "Exemplary Unpopular Opinion" is attached to the right side of the title. It remarks that this post is an outstanding unpopular opinion post. Therefore, users could use the flairs to pick out only the well written unpopular opinions in r/unpopularopinion in the past with the old design.

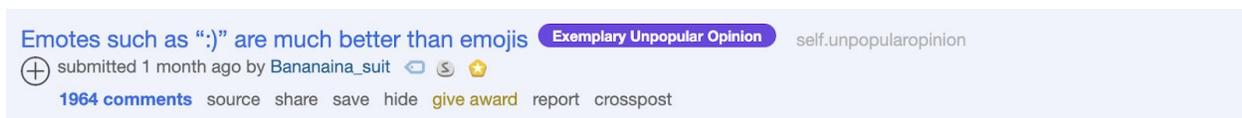

*Figure 8.* An "exemplary unpopular opinion" flair

*Sorting Option*

The discussion board of r/unpopularopinion was designed to show all the discussion threads in one page. The threads can be sorted by five algorithms: *hot*, *new, controversial, top,* and *rising* (Figure 9).

- *Hot* show the threads which are recent and highly upvoted first;
- *New* sorts the post by the most recent posts created by the users;
- *Controversial* shows threads that have a roughly equal number of upvotes and downvotes, which aggregate a high number of votes;
- *Top* sorts vote by the number of upvotes in a particular time period;
- *Rising* shows the threads which are gaining high upvote momentum first;

In r/unpopularopinion, users leave comments on each post thread. The comments on the post are also filed up just as like the posts do. The similar sorting algorithms are applied to comments threads with two additional algorithms: sort by oldest and sort by Q&A, which shows only the comments which the original poster replies to.

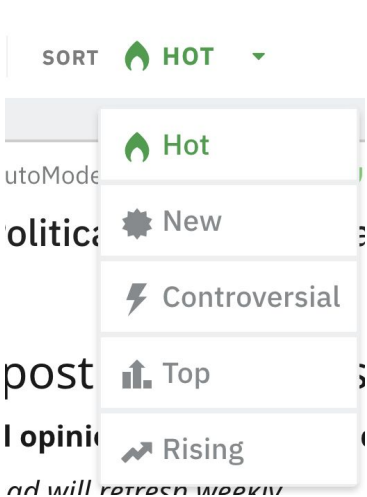
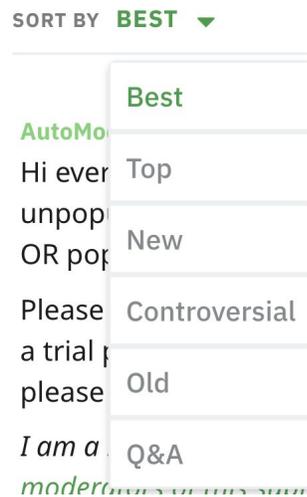

*Figure 9*. The drop-down menu of sorting option for posts in r/unpopularopinion

*Figure 10*. The drop-down menu of sorting option for comments in r/unpopularopinion

**Distinct and Extensible Gathering Places**

In r/unpopularopinion, users fulfill their needs to have a third place in which they can take a rest and talk about any topics without any backlash towards their real-world selves. Users are usually protected by their alter-ego in the cyberspace, and it is hard for the others to match the user in r/unpopularopinion to the user's actual self. Thus, users in r/unpopularopinion do not need to worry about the risk of their privacy to be invaded and their identity to be revealed [8]. It provides users with a comfortable place to freely express their feelings and opinions on any topic.

**Membership**

*Barriers to entry and Benefits of membership*

Anyone with an internet connection can access the r/unpopularopinion and read any posts and comments in the community. However, to be able to write and actively participate in a discussion, people need to create a Reddit account since r/unpopularopinion is one of the subreddits, sub-discussion board, of Reddit.  The benefits of creating a Reddit account include the ability to subscribe, vote and comment on the Reddit content (Figure 11). After joining Reddit, there is no barrier to subscribe to r/unpopularopinion, but users with a new account need to wait 24 hours to be able to post in r/unpopularopinion; this feature is used as a spam prevention measure in Reddit.

*Pseudonymity*

Donath presented an idea that each user has an online pseudonym, which is decoupled from the real identity [4]. The Reddit community embraces this idea while opposing doxing, which means exposing personal information. Reddit users do not need to disclose anything about their personal information to participate in Reddit activity. Instead, they can choose a pseudonym and interact with other members with their adopted pseudo-identity. Reddit users can have multiple accounts for different purposes and contexts. One example is Participant 4, one of the interview participants, who has two accounts for moderating and browsing.

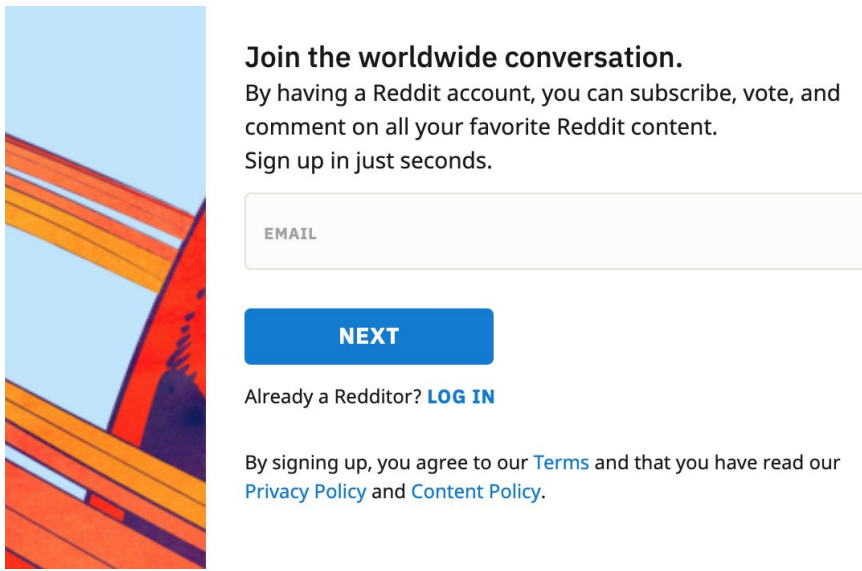

*Figure 11.* The sign-up page of Reddit explains the benefits of having a Reddit account.

*Evolving Member Profiles and Personal Profile Page*

Each Reddit user has a profile page. The user's profile on the right tab of the page shows the number of karmas, which represents the number of upvotes the user has received since the Cake day, which is the day the user created the Reddit account. In the profile page, the user can freely edit the profile description to introduce himself to other users or add posts on this page just as like it can be done in other subreddits. Reddit users typically choose to not reveal their real identity; however, they can choose to disclose their real identities on their profiles if they want. Also, the profile page shows other information like frequently interacted subreddits, subreddits which users moderate, and award badges. Moreover, the profile page shows the history of posts and comments made by the user, which cannot be hidden.

*Membership status as 'flair'*

In the subreddit r/unpopularopinion with the old design, users could use the 'flair' to represent his membership status (Figure 12). This feature combined with the regular Reddit profile creates a pseudonym for a Reddit user.

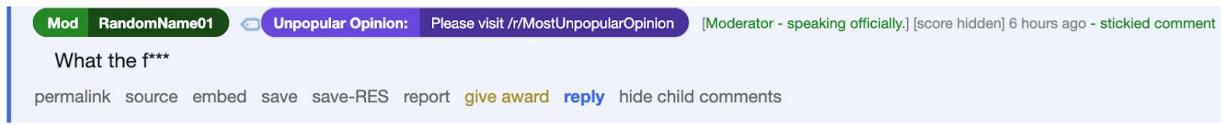

*Figure 12.* A flair indicating that the user who posted this comment is a moderator of r/unpopularopinion.

**Range of roles**

There are three major roles in r/unpopularopinion: visitor, member, moderator. The first role is a visitor, who can read all the posts and comments on the subreddit. The visitor can be considered as a lurker in a point that they only consume the contents of the community, but not contribute to the community. The next role is a member, who can also create a post and leave a comment on other posts. Every post and comment is upvoted or downvoted by members, and the rank determined from the votes moves the post to front or back page based on the total score of the post in the community. Not only the regular users of the subreddit r/unpopularopinion, but all the other users of Reddit can participate in voting on any posts or comments of the subreddit r/unpopularopinion.

The r/unpopularopinion has a *moderator* team, which is chosen from the pool of *member*s. The *moderators* have the authority to change the visual design of the r/unpopularopinion page and to add more functionalities by creating custom CSS and Reddit bots. They are responsible for reviewing the comment and post contents and remove them if necessary. They also hand out punishments to users who violate the rules of the subreddit. The *moderator* posts and comments are shown as "announcement from the moderators". Also, the moderators are responsible for posting megathreads, which serve as the discussion board for politics and other hot issues.

*Promoting effective leadership*

The moderators are chosen from the user base in a recruiting post on r/unpopularopinion. Participant 20 shared his recruiting experience.

> "I found Unique (another moderator). He maybe messaged me like four years ago or five years ago when the sub was still pretty small. And then we decided, or we found a couple other moderators over that time and none of them really worked out. And then about eight months or so ago, he made a Google, like a Google sheet and then put up a post recruiting moderators. And then we just picked out the top, you know, six or so that he liked the best and we interviewed them, and they've all worked out great. So now we have a really good team."

It is common in Reddit to have volunteers from the user base to become a moderator. As the moderators are volunteers, they may not actively manage the community. Participant 4 shared his experience as a moderator that the moderation team was not active in enforcing the rule one year ago, and they had to recruit other new moderators consequently. The newly recruited moderators were much more active in managing r/unpopularopinion, and therefore r/unpopularopinion has grown and the number of subscribers has increased greatly.

**CODE OF CONDUCT**

The main focus of the subreddit moderation team is building a community that facilitates civil and interesting discussions. It implements a strong policy against trolling and hate posts. Members who troll or write hate posts and comments are permanently banned. Likewise, spamming leads to permanent bans and new accounts have to wait 24 hours before posting. Members are encouraged to report the posts which violate the above rules. The reported posts by users or auto-moderator are reviewed again by the human moderators. Then the moderators decide whether to ban the owner of the posts. The moderators have a Discord channel to discuss violations cases, which are typically about 10 cases a day. Sometimes, an individual moderator can make a decision by at his or her own discretion on the violation, which had lead to a funny situation in the past that a moderator banned a second account of another moderator. Later, the moderator whose second account was banned unbanned himself.

In r/unpopularopinion, members are expected to upvote the unpopular/controversial and well-written opinions and downvote popular or poorly written opinions. The definitions of unpopular opinions are not specifically defined by the moderation team. Therefore, each member needs to decide whether an opinion is unpopular based on their own environments, background and life experience. This is one of the most interesting points of r/unpopularopinion, and it will be expanded more upon in **Research questions**.

The r/unpopularopinion has a strict guideline for writing an opinion. A post must have a self-contained opinion within the title. This rule was set to prevent users from having to open the post to read what the claim is and to make the subreddit page look more neatly and more interesting. The text of post content should be used to elaborate and explain the owner's opinion. It has to be longer than 160 characters; otherwise, it will be deleted by the AutoMod.

To prevent reposts, users are encouraged to search through the previous posts on r/unpopularopinion before posting an opinion to check if similar opinions have been posted previously. Commonly posted opinions in r/unpopularopinion are not considered a valid unpopular opinion and may be deleted by the Auto-moderator if they contain certain excessively common keywords. Similarly, posts with low efforts or satire posts are not allowed, as well as opinions about obscure topics.

Recently, a new rule was added to explicitly ban two types of posts: Meta posts, which are posts about the current states of r/unpopularopinion, and posts about politics. Participant 4 said "[Before] one in every three posts will be about the Subreddit itself. And then the other one is US politics". If users violate these rules, then they will be punished with a temporary ban for 24 hours, as a warning. Participant 4 further mentioned that the Auto-moderating system was tweaked extensively to remove these types of posts as a result of adding these rules.

As a result of the recent changes, the moderators received many foul messages. On removing a political post, Participant 4 shared his experience:

> " ... We just straight up remove [the political threads]. And we'll get messages from a user saying: "Oh my God, you conservative c**ks, how dare you remove my Trump post?" Or my Obama post or whatever. "

There is currently a trial period in which all political opinions are put into a megathread, which is detailed more in **CYCLIC EVENTS**.

Similarly to other subreddits, the rule amendment is mainly discussed in the moderators' private Discord channel. Currently, Two moderators are responsible for implementing the Auto-moderating system. Because the Auto-moderator implementation cannot be openly disclosed to block any loophole to be used by trolls, the members are often confused with random legitimate posts being deleted by auto-moderator by mistake. To improve the auto-moderation in r/unpopularopinion, *Design claim 22*, which advocates community influence on rulemaking, could be applied to make the moderation process more democratic [9].

**CYCLIC EVENTS**

*Weekly Political Opinion Megathread*

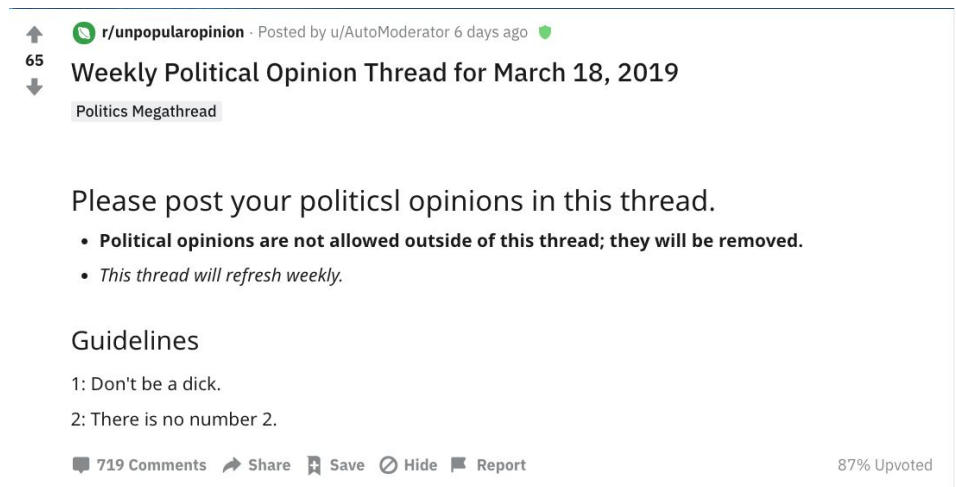

*Figure 13.* Weekly political opinion thread of r/unpopularopinion.

The subreddit r/unpopularopinion has a weekly political opinion megathread. For most of the time, it is pinned on top of the front page of this subreddit (Figure 13). As it was mentioned above, the tenth rule of this community is "No US Politics" in r/unpopularopinion. The moderators make it clear that there are a number of other places to discuss political issues in Reddit and r/unpopularopinion is not one of them. However, there are still many people who want to discuss political issues in this forum, so the moderators confine all the political issues to be debated only in this weekly megathread. Having a weekly political opinion megathread is a very controversial issue in r/unpopularopinion since political discussions are often overheated, and the users tend to discuss based on a dichotomous view of Republicans versus Democratic. Some users find it interesting to discuss political issues in r/unpopularopinion. One participant mentioned that users should have a right to post political opinions in this subreddit if they want:

" I think Reddit at its core, I really like the philosophy of its self-governing. If people want to post that stuff, that's fine, let them post it. If people get sick of it, they'll just

downvote it, and it won't make it to the top pages."

Other users do not want to see the political topics in this subreddit but want other users to discuss politics in subreddit politics. One interview participant shared his opinion on this political opinion megathread as follows:

"I honestly tend to stay out of that one. Honestly, just because I think that yelling at

people to politics. There's never going to be a real solution. I don't think anyone's really

going to change their minds."

Tons of political issues are regularly produced every week, and users in this weekly politics megathread discuss whether they support it or not. A lot of comments are not likely to be considered unpopular opinions in this weekly megathread, but users are actively participating in the discussion in it.

*Megathread: New Zealand Shootings*

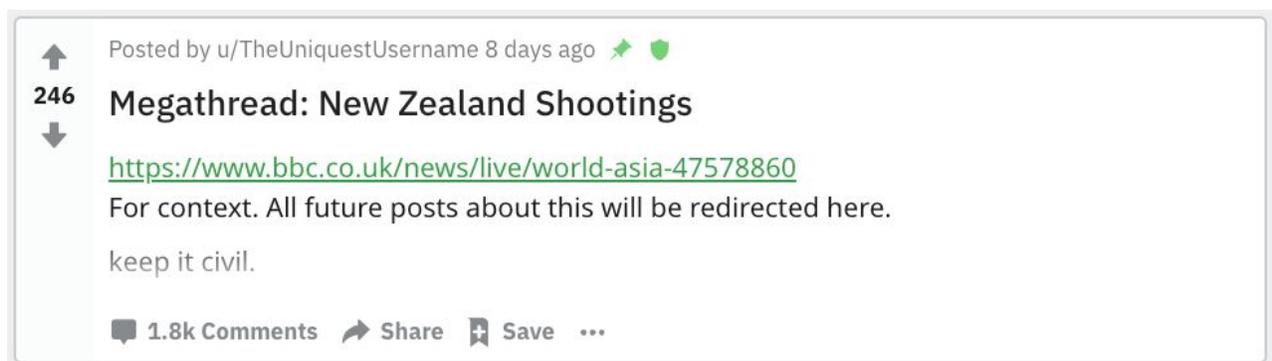

*Figure 14.* Megathread: New Zealand shootings in r/unpopularopinion.

When there is any recent debatable worldwide issue, and a lot of users post their opinions on the same issue repeatedly, the moderators of r/unpopularopinion start a megathread and pin it on the very top of the front page so that people can keep discussing the issue in one spot. The moderators want to keep the discussion board of r/unpopularopinion to present diverse opinions on various topics but not to be occupied by one topic. Recently, on March 17, 2019, the moderators created a new megathread about the New Zealand mosque attack which occurred on March 15, 2019 (Figure 14). The moderators clearly stated on top of the megathread that all the related posts will be redirected to this megathread.

**Research questions**

For the community analysis, we decide to focus on four research questions:

1. What do subreddit unpopularopinion users perceive as unpopular opinions?
2. What does r/unpopularopinion have to help people break the spiral of silence?
3. Voting works differently in r/unpopularopinion comparing to the rest of Reddit. How does it shape the subreddit?
4. Is there an echo chamber effect on this subreddit?

*Unpopular opinion*

In this section, we discuss what the unpopular opinion is and how the interview participants think about the unpopular opinion in r/unpopularopinion.

    **The US President Donald Trump** The definition of an unpopular opinion varies across different members of the subreddit r/unpopularopinion. It is still a fog as to what makes an opinion actually unpopular. In the interview, one of r/unpopularopinion moderators mentioned:

> "Is disliking Donald Trump an unpopular opinion? I do not think so. He won the election. He cannot be that unpopular, but it is still controversial, you know?"

This is a good example that the moderator gave to define what unpopular opinion is. Before the 2016 United States presidential election, mainstream media had condemned him for the actions he had taken, and a lot of people openly criticized him. Likewise, most media predicted that he would be defeated by a large margin in the presidential election. Therefore, Donald Trump had been considered as a generally disliked figure. However, he won the presidential election with the support of many US citizens surprisingly, and this phenomenon posed a question about what the popular opinion is in modern society. Mass media constantly criticized Donald Trump because this attracted a lot of viewers and readers. Have we been brainwashed with specific information by the mass media for their benefits? Because most of the press said that Donald Trump had no chance to win the election, the public was shocked when he was elected. Later, the media interpreted that nonvocal Trump voters supported him, but we do not know if it is true or not. In fact, the only information the public knows about Donald Trump is the information provided by the press. What else do we know about him personally? In modern society, the public trusts the information provided by the media, decide and act based on the information. However, if the media intentionally distorted the information, the public could be used and played by the media. Is liking Donald Trump, who won the presidential election in the United States, still an unpopular opinion? Why Trump's supporters keep silent? The press nurses the public atmosphere and stipulates what popular opinion is. If the press only published favorable articles about Trump, the public would have supported him regardless of whether he is a good president or not. Therefore, it can be concluded that public opinions are created by the media in modern society.

**Popular opinion vs Unpopular opinion** In the other interviews, many participants defined that an unpopular opinion is generally unappealing to the majority of the population. At this point, how do you know if the majority of the public agree or disagree with the opinion? Moreover, if the opinion is not favorably accepted by mainstream society, does it make the opinion unpopular? "I do not hate it." does not mean "I like it.". Similarly, the fact that it is not a popular opinion does not mean that it is an unpopular opinion. To make an opinion considered popular or unpopular, the vast majority of people should have a common opinion on it. For example, almost all of the people on earth agree with the fact that murder is a serious crime, and it is a popular opinion. If 60% of the population agrees with the opinion while the other 40% do not agree with it, then the issue can be considered rather as a controversial subject.

**Unique ideas** One interview participant gave another idea regarding what an unpopular opinion is. He mentioned as follows:

> "... I think that if you look at the top posts of all time, a lot of those follow that spirit where they are a unique viewpoint, rather than necessarily a completely unpopular opinion"

For him, an unpopular opinion was the thought from a unique perspective. Considering the definition of unpopular opinion from other participants that unpopular opinion is not appealing or unsavory to the mainstream society, a unique viewpoint might not necessarily be unappealing to everybody. Yet, because it is a unique opinion, it makes sense that it can fall into the range of unpopular opinion.

**Context** Now, participant 5 of the subreddit gives a whole new perspective to what is considered as an unpopular opinion in r/unpopularopinion. She mentioned in the interview that whether an opinion is unpopular or popular depends on the audience. If audiences agree with the opinion, then it can be a popular opinion among their group, even if it is not accepted as popular opinion in the mainstream society. Moreover, the opinions popular opinion in Asian countries might be unpopular in Europe and vice versa. That is, context and the audience in the conversation may influence how the opinion is accepted.

*Breaking the spiral of silence*

Noelle-Neumann theorized that people have a fear of being isolated if they think differently from the rest of their social group; therefore, people tend to remain silent instead of speaking out their frank opinions to hide their different idea [14]. The community design of r/unpopularopinion has provided people some ways to overcome the spiral of silence.

**Pseudonymity**. Reddit is run based on the pseudonym user system. The Reddit user account is not related to anything from the real-life identity of the user. Therefore, the user does not have to fear to be neglected from their real-life social groups by speaking out their opinion in Reddit activity. Most interview participants stated that they do not want to share their account activities with their families, close friends or employers. Participant 4 stated that:

> "God, no. That would be really bad. That'd be really, really bad. My post history on my mod account is probably fine. I am not sure. I would be surprised if it was not. But my post history on my alternate account is not good."

In *The presentation of self in everyday life*, Goffman wrote that humans play different performances based on the current social roles and contexts [6]. Similarly, the user's behavior in r/unpopularopinion can be different from their real life. Also, the user can have many Reddit accounts and play different roles carrying different perspective to fulfill different purposes. For example, moderators in r/unpopularopinion typically use two accounts for moderating the community and participating in regular user activity. Participant 4 stated:

> "Usually when I am on my mod account, I mostly just interact with ideas that I think are poorly thought out. Because normally what I am doing on my mod account is just reviewing the latest post to make sure everything is cool. And if a post does not break the rules, but I still think it is kind of stupid, I might comment on it and criticize it, but on my main account, on my other account, that none of the other mods even know about.

…

On my alternative account, I post a lot of very, very controversial things. In fact, I have had to unban myself at least once on my mod account because somebody on the mod team took it a little harsh, the posts I might have put on that account. So I post some very unpopular things and it kind of got to the point where there were a bunch of people that were almost out to get me."

**Controversial contents are encouraged**. Yun and Park suggested that "in an online forum where a majority opinion exists, users may decide not to post a message against the majority opinion." [18]. In contrast, upvoting on well-written unpopular opinions is encouraged in r/unpopularopinion. This community policy attracts many people who are interested in controversial views and discussions. Participant 5 told us:

"I think there is a morbid curiosity in hearing controversial and dark views. People tend to be interested in controversial things and there is plenty of that on r/unpopularopinion."

Therefore, there is an incentive in the form of Reddit Karma score for users to speak out against the perceived climate. The higher the karma, the higher the attention the original poster gains. On top of that, the moderators implement a neutral and free speech policy. Adam shared:

"Yes, people will often disagree with an opinion posted there but will actually upvote the post for being an unpopular opinion. Like I said earlier the mods and users are all very pro-free-speech so you can have a quality discussion on that sub."

**Civil discussions are encouraged.** While there is a common complaint that some unpopular opinions are not upvoted enough, the subreddit's *sort by controversial* feature means that unpopular opinions can also receive a lot of attention through comments. The policy of r/unpopularopinion states that unpopular opinions need to be elaborated up to a certain length so that other subreddit users can understand exactly what the original poster's opinion and discuss around it. Participant 4 shared his opinion:

> "I would say yes and no. I would say that it is a lot more open to having discussions about very controversial topics. But at the same time, people do get kind of a little bit angry on the Subreddit because a lot of the issues are very important to them. So, there is a lot of name calling and that kind of thing and that has to be taken care of. But I would say that overall, people who can say things that are not very okay in Reddit and regular society and still have people civilly discuss them, which I like."

The attention received from a post can be an incentive for users to speak against the opinions of the majority.

*Voting behavior in r/unpopularopinion*

**Voting design in r/unpopularopinion** The way the voting is performed in r/unpopularopinion is different from the Reddit norms: r/unpopularopinion encourages users to upvote on opinions they consider unpopular or controversial. The moderation team of r/unpopularopinion agrees that the users should decide arbitrarily what is popular/unpopular. There are two reasons for this: each people's concept of unpopular opinion is different and there are no moderation tools available for regulating voting behavior of each user. A moderator shared his thought:

> "Because upvotes and downvotes are completely anonymous, it is kind of impossible to have a rule that enforces something that's inherently unenforceable. We cannot enforce people's upvoting or downvoting because we cannot see it. So it is really more, this is the way it should be."

Some interviewees were aware of the proper way of voting in r/unpopularopinion. Astro said:

> "I stick with the guideline. If I see something that I disagree with, but it is somewhat well written and not just a personal attack or just outright hatred, then I upvote."

As a result, the more controversial and unpopular an opinion is, the more likely it goes to the front page. Participant 4 shared his first impression of the subreddit.

> "It was a sort of free space to talk about whatever you like. And I joined just because a lot of people post interesting things and I do not know if you use Reddit regularly, but the front page of Reddit or the more popular Subreddits are very, I do not want to say clinical exactly, but they kind of rehash the same ideas in a very family friendly way. And it's very difficult to discuss anything serious or controversial on those Subreddits."

Ling and his colleagues confirmed the hypothesis that there were more user interactions in a group of people who disagree with each other than in a group of people who agree with each other [11]. Similarly, in subreddit r/unpopularopinion, controversial topics drive more user interaction. Ketogamer said:

> "The typical [opinion] I see is usually like an anti-woman thing. I love commenting on those. Jokes are one thing, but I'll see tons of unironic, serious comments where they're like, "You know, women, they are just too emotional and they are stupid for their own good." I'll be like, "I do not know what you are talking about. Do you have any proof?" They're like, "Come on, it is obvious." I probably spend too much of my time arguing about that."

However, many users had the impression that r/unpopularopinion has many popular opinions on the front page while unpopular opinions are not upvoted enough.

**Popular opinions are upvoted, but not downvoted.** All of the users we interview acknowledges that sometimes a popular opinion in r/unpopularopinion is upvoted and appear on the front page. A possible reason for this behavior is that users in r/unpopularopinion do not follow the encouraged voting behavior. Astro shared his experience in writing a meta post.

> "A while ago I posted a meta post about this Subreddit essentially being like what I said, they aren't necessarily unpopular. They are just socially less accepted opinions that are posted here. Because if you get upvoted, it means that your opinion is popular. Even people on this sub are still not strong enough or I guess impartial enough to defeat their tendency to just downvote something that they disagree with."

Ketogamer shared his thought:

> "People are not upvoting what is truly an unpopular opinion. They are upvoting with an opinion that they agree with, that they think is unpopular from the rest of the world, but that they like it themselves."

This is quite likely, as the encouraged voting behavior in r/unpopularopinion is somewhat different from the rest of Reddit. The *hot* page is the front page of each subreddit. The more upvote a post receives, the more likely the post goes to the front page. Any experienced Reddit users know this feature; therefore, they generally upvote the content they want to see more and downvote the posts that they do not want to see. Likewise, the users usually do not downvote on popular opinion if seeing the post does not make the user angry but fun.

Some Redditors bring their general voting behavior to r/unpopularopinion. While Astro is aware of the encouraged voting behavior in r/unpopularopinion, he does not downvote a popular opinion. Participant 6 was not aware of the voting rule of r/unpopularopinion.

> "Usually yeah, I tend to [upvote] on the ones that I agree with but like I said, there's not many but I usually downvote once I do not agree with it, but I mean I try to remain relatively impartial and get my opinions on those that I do and do not agree with."

To make it worse, a huge number of users use the new Reddit redesign and Reddit Mobile App, which do not show the tooltip message, which reminds the users of the voting rules in r/unpopularopinion. As the new design was improved, the moderators recently added an AutoMod message to pops up on the top of each thread to reminds the voting rules, and this change has solved this problem to some degree.

Voting ability is also not limited to r/unpopularopinion users. Especially, when a post from r/unpopularopinion receives a lot of upvotes and appears in the r/all, the front page of Reddit, the post is exposed to Redditors who are not r/unpopularopinion members. Since they do not know the voting rule of r/unpopularopinion properly, they vote on the content they agree with as they do in other subreddits. Therefore, *top* posts of r/unpopularopinion tend to have more "popular" opinions.

Another likely reason is that each people's standard on what is an unpopular opinion: one person might consider it as a popular opinion while the other considers it an unpopular opinion as discussed above in the paper.

**Downvote on unpopular opinions** Encouraged voting behavior in r/unpopularopinion can be emotionally taxing for users as follows. Reddit voting mechanism is partially designed to keep out extreme and offensive views. Sorting by controversial in r/unpopularopinion can be an unpleasurable experience for some users. Ketogamer shared his opinion:

> "I think, on the whole, it is good at handling the people who are just saying something so out of left field that it just has no place for being there. If someone is saying, "Hey, unpopular opinion, we should just gather up a bunch of children and just murder them for fun." If it is something so stupid and so ridiculous, I think the downvotes do a good job of keeping that away."

Many unpopular opinions can be highly offensive to other users. Accordingly, they do not get a lot of upvotes. Also, there are many other reasons that an unpopular opinion is not getting upvoted: timing, poorly written content, obscure topics, topic interest levels. This can frustrate users who expect interactions and attention from posting an unpopular opinion. This can create an impression that unpopular opinions are not upvoted in r/unpopularopinion.

**Partial Ephemerality** Reddit encourages users to keep voting on new posts on a continuous basis. A lot of posts disappear from the front page, and even the most popular opinions do not stay on the front page for more than one to two days. Once the opinion disappears from the frontest page, it is very hard to find the previous post with the current sorting options, but the search should be done with a brute-force search method. Therefore, if a user does not access r/unpopularopinion for a day, he cannot know what kind of opinions were discussed on the day. Consequently, it is difficult for users to know the past discussion trends in r/unpopularopinion, and it makes people post opinions that might have been uploaded before.

**The repercussion of Similar opinions** If a user encounters similar opinions on a certain topic repeatedly in r/unpopularopinion, he may receive an impression that popular opinions keep getting upvoted again and again.

*Echo chamber*

In the participant interview, one of the participants said that the comments section of each post often shows the echo chamber effect. He mentioned as follows:

"... It is odd because you would expect unpopular opinions to get backlash, but then the posts result in an echo chamber of so many people having these so-called "unpopular" opinions."

Because the nature of r/unpopularopinion is for users to post unpopular or controversial opinions, it is very interesting that a lot of users leave a comment that they agree with the posted opinion, which is supposed to be unpopular. One of the interview participant who is a moderator of r/unpopularopinion mentioned that even the posts show the echo chamber as there are many posts covering the same opinion but worded differently or just recycled. Moreover, reposting is a popular way to discuss an old discussion topic, it contributes to the echo chamber in r/unpopularopinion as well.

One participant mentioned that this phenomenon might come from the fact that r/unpopularopinion is mostly populated with young white males, and they might share similar opinions only in their group which might be unpopular to the mainstream society. In r/unpopularopinion, the repeatedly uploaded similar posts and comments result in the echo chamber where the young white male's beliefs and opinions are reinforced toward a more extreme perspective.

**Conclusion**

*The spiral of silence in r/unpopularopinion*

Yun and Park presented quantitative evidence that "people's perceptions about their position as being in the majority or the minority in the cyberspace can reduce their willingness to speak" [18]. The partial ephemerality in r/unpopularopinion causes users to take some effort and time to perceive their position being in the majority or the minority in r/unpopularopinion. However, once the user realizes that their opinion is within the minority and is not welcomed by other members of r/unpopularopinion, they become hesitant to speak out their opinion in r/unpopularopinion; which reveals that the phenomenon of the spiral of silence is repeated among the members of r/unpopularopinion.

**Implications**

In the future study, we can further discuss the user behaviors against the spiral of silence in an anonymous online community like 4chan. Would the users in the anonymous community show a stronger reaction against social conformity and post more frank opinions?

One of the problems of r/unpopularopinion we have found in the study is that similar opinions on the same topic are reuploaded repeatedly. After being exposed to similar unpopular opinions continuously, users may perceive the idea to be popular and general. To solve the re-post issue, the community can introduce the filter by category feature and the topic tags, or delete the newer post and link the original post.

 Moreover, the voting system of r/unpopularopinion comes directly from the Reddit voting system in which the popular opinions to be more frequently exposed to the users. In contrast, r/unpopularopinion is designed for users to upvote the post of unpopular opinion, and thus a lot of users are confused about the way to upvote or downvote in the current r/unpopularopinion community. Even, the community rules section on the right tab of the r/unpopularopinion page does not clearly state this function. Therefore, we suggest the community to clearly represent the proper way of voting in the community rules.

In addition, there are too many opinions posted for moderators to check and manage in r/unpopularopinion. Consequently, automoderator manages the posts of r/unpopularopinion by searching specific problematic keywords and deleting the post or banning the user. However, the current auto-moderation does not cover all cases properly; some good posts are wrongly deleted while some bad posts are not deleted properly. To better manage r/unpopularopinion, more human moderators are needed until a better auto-moderating system is developed.

Finally, the current r/unpopularopinion is representing opinions of the specific population, which is assumed to be young male adults, and this community does not incorporate diverse user groups. As a result, opinions speaking for the major group, young male adults, in r/unpopularopinion get the support of users while opinions representing the minority in r/unpopularopinion are neglected. To embrace more diverse opinions of users from many different backgrounds, r/unpopularopinion needs better rules to make users to respect different opinions [8]. The new question is: How are we going to make people respect different thoughts and other people in an online community?

SUBREDDIT R/UNPOPULAROPINION                                                                                           42

References

[1] Bikhchandani, S., Hirshleifer, D., & Welch, I. (1998). Learning from the behavior of others: Conformity, fads, and informational cascades. *Journal of economic perspectives*, *12*(3), 151-170.

[2] Chaudhry, I., & Gruzd, A. (2019). Expressing and Challenging Racist Discourse on Facebook: How Social Media Weaken the "Spiral of Silence" Theory. *Policy & Internet*.

[3] Cook, K. S., Cheshire, C., Rice, E. R., & Nakagawa, S. (2013). Social exchange theory. In *Handbook of social psychology*(pp. 61-88). Springer, Dordrecht.

[4] Donath, J. S. (2017, June 03). We Need Online Alter Egos Now More Than Ever. Retrieved from https://www.wired.com/2014/04/why-we-need-online-alter-egos-now-more-than-eve/

[5] Friedkin, N. E., & Johnsen, E. C. (1990). Social influence and opinions. *Journal of Mathematical Sociology*, *15*(3-4), 193-206.

[6] Goffman, E. (1978). The presentation of self in everyday life (p. 56). London: Harmondsworth.

[7] Hampton, K. N., Rainie, H., Lu, W., Dwyer, M., Shin, I., & Purcell, K. (2014). *Social media and the 'spiral of silence'.* PewResearchCenter.

[8] Jhaver, S., Vora, P., & Bruckman, A. (2017). Designing for Civil Conversations: Lessons Learned from ChangeMyView. *Georgia Institute of Technology*.

[9] Kiesler, S., Kraut, R., Resnick, P., & Kittur, A. (2012). Regulating behavior in online communities. *Building Successful Online Communities: Evidence-Based Social Design. MIT Press, Cambridge, MA*.

[10] Kim, A. J. (2000). Community building on the Web. *Berkeley, Calif: Peachpit Press*.

[11] Ling, K., Beenen, G., Ludford, P., Wang, X., Chang, K., Li, X., ... & Resnick, P. (2005). Using social psychology to motivate contributions to online communities. *Journal of Computer‑Mediated*

SUBREDDIT R/UNPOPULAROPINION                                                                                43

*Communication, 10*(4), 00-00.

[12] Liu, X., & Fahmy, S. (2011). Exploring the spiral of silence in the virtual world: lndividuals willingness to express personal opinions in online versus offline settings. *Journal of Media and Communication Studies*, *3*(2), 45-57.

[13] Malaspina, C. (2014). The spiral of silence and social media: Analysing Noelle-Neumann's phenomenon application on the Web during the Italian Political Elections of 2013. *London School of Economics and Political Science*.

[14] Noelle-Neumann, E. (1984). The spiral of silence (pp. 6-7). *Chicago: University of Chicago Press*.

[15] Phelan, J. C., Link, B. G., & Dovidio, J. F. (2008). Stigma and prejudice: One animal or two?. *Social science & medicine*, *67*(3), 358-367.

[16] R/unpopularopinion - r/UnpopularOpinion Demographics Survey - RESULTS ARE IN! (n.d.). Retrieved from https://www.reddit.com/r/unpopularopinion/comments/9opad4/runpopularopinion_demographics_survey_results_are/

[17] Subreddit Stats. (n.d.). Retrieved from https://subredditstats.com/r/unpopularopinion http://r/unpopularopinion.%20(n.d.).%20Retrieved%20from%20https://www.reddit.com/r/unpopularopinion/

[18] Yun, G., & Park, S. Y. (2011). Selective posting: Willingness to post a message online. *Journal of Computer-Mediated Communication, 16*(2), 201-227.

APPENDIX A

Informed Consent Form

https://gatech.co1.qualtrics.com/jfe/form/SV_bdt8UiZNWiEf5sx